\documentclass[useAMS,usenatbib,usegraphicx]{mn2e}

\title[Probing the behaviour of Cyg X-3 with VLBI]{Probing the behaviour of the X-ray binary Cygnus X-3 with very-long-baseline radio interferometry}
\author[V. Tudose et al.]{V. Tudose,$^{1,2,3,4}$\thanks{E-mail: tudose@astron.nl (VT)} J.C.A. Miller-Jones,$^{5}$\thanks{Jansky Fellow, National Radio Astronomy Observatory} R.P. Fender,$^{6,2}$ 
Z. Paragi,$^{7,8}$ C. Sakari,$^{9}$ \newauthor A. Szostek,$^{10,11}$
M.A. Garrett,$^{1}$ V. Dhawan,$^{12}$ A. Rushton,$^{13}$ R.E. Spencer$^{13}$
\newauthor and  M. van der Klis$^{1}$ \\
$^{1}$Netherlands Institute for Radio Astronomy, Postbus 2, 7990 AA Dwingeloo, the Netherlands\\
$^{2}$Astronomical Institute `Anton Pannekoek', University of Amsterdam, Kruislaan 403, 1098 SJ Amsterdam, the Netherlands\\
$^{3}$Astronomical Institute of the Romanian Academy, Cutitul de Argint 5, RO-040557 Bucharest, Romania\\
$^{4}$Research Center for Atomic Physics and Astrophysics, Atomistilor 405, RO-077125 Bucharest, Romania\\
$^{5}$National Radio Astronomy Observatory, 520 Edgemont Road, VA 22903 Charlottesville, USA\\
$^{6}$School of Physics and Astronomy, University of Southampton, Highfield, Southampton SO17 1BJ\\
$^{7}$Joint Institute for VLBI in Europe, Postbus 2, 7990 AA Dwingeloo,
the Netherlands\\
$^{8}$MTA Research Group for Physical Geodesy and Geodynamics, P.O. Box
91, H-1521 Budapest, Hungary\\
$^{9}$Whitman College, 345 Boyer Ave., WA 99362 Walla Walla, USA\\
$^{10}$Laboratoire d'Astrophysique de Grenoble, UMR 5571 CNRS, Universit\'e Joseph Fourier, BP 53, 38041 Grenoble, France\\
$^{11}$Astronomical Observatory, Jagiellonian University, Orla 171, 30-244 Krak\'ow, Poland\\
$^{12}$National Radio Astronomy Observatory, 1003 Lopezville Road, NM 87801 Socorro, USA\\
$^{13}$Jodrell Bank Centre for Astrophysics, School of Physics and Astronomy, University of Manchester, Manchester M13 9PL}

\begin{document}

\date{Accepted 2009 September 15. Received 2009 September 14; in original form 2009 July 16}

\pagerange{\pageref{firstpage}--\pageref{lastpage}} \pubyear{2009}

\maketitle

\label{firstpage}

\begin{abstract}
\noindent
In order to test the recently proposed classification of the radio/X-ray states of the X-ray binary Cyg X-3, we present an analysis of the radio data available for the 
system at much higher spatial resolutions than used for defining the states. The radio data set consists of archival Very Long Baseline
Array data at 5 or 15 GHz and new electronic European Very Long
Baseline Interferometry Network data at 5 GHz. We also present 5 GHz Multi-Element Radio
Linked Interferometer Network observations of an outburst of Cyg X-3. In the X-ray regime we
use quasi-simultaneous with radio, monitoring and pointed Rossi 
X-ray Timing Explorer observations. We find that when the radio emission from both jet and core is globally considered, the behaviour of Cyg X-3 at milliarcsecond 
scales is consistent with that described at arcsecond scales. However,
when the radio emission is disentangled in a core component and a jet
component the situation changes. It becomes clear that in active states
the radio emission from the jet is dominating that from the core. This
shows that in these states the overall radio flux cannot be used as a direct
tracer of the accretion state.

\end{abstract}


\section{Introduction}

Cygnus X-3 (Cyg X-3) is an exotic X-ray binary system (XRB) discovered in the X-ray band by \cite{Gia67}. The nature of the compact object is unknown, 
circumstantial evidence existing for both a black hole \citep*{Che94,Sch96,Hja08,Szo08a} and a neutron star \citep*{Tav89,Mit98,Sta03}. Strong 
evidence points toward a Wolf-Rayet star companion \citep*{Ker96,Fen99,Koc02} which makes Cyg X-3 a rather special object among the XRB population 
given the fact that at present only two other objects of this class, both extragalactic \citep{Pre07,Car07}, seem to harbour such a companion star type. 
The system is at a distance of 7--9 kpc \citep*{Pre00,Lin09} and has an orbital period of 4.8 h inferred from infrared (e.g. \citealt{Bec73}) and X-ray 
(e.g. \citealt{Par72}) observations. 

In the radio band the system shows flares of different amplitudes, the strongest of them reaching up to a few tens of Jy at cm wavelengths 
(e.g. \citealt*{Wal94,Tru08}). During these outbursts, Cyg X-3 reveals the presence of relativistic jets \citep*{Gel83,Spe86,Mol88,Sch95,Sch98}, with a 
complex structure that was clearly resolved in a few occasions \citep*{Mio01,Mar01,Mil04,Tud07}. Based on arcsec-scale radio observations 
\cite{Wal94,Wal95,Wal96} identified four distinct radio states of the system: quiescent (flux densities $\sim$ 100 mJy), minor flaring ($<$ 1 Jy), 
major flaring ($>$ 1 Jy), and quenched ($<$ 30 mJy). 

In the X-ray band the study of Cyg X-3 has been hindered by the strong absorption existing in the system which is likely caused by the wind of the companion star 
\citep{Szo08a}. The system shows two main X-ray states, hard and soft, plus a couple of transitional ones that are less understood (e.g. \citealt{Hja09}). These 
states are similar to the canonical X-ray states of black hole XRBs (e.g. \citealt{Zdz04}).

Quasi-simultaneous multi-wavelength studies, focused primarily on the X-ray and radio bands, have been carried out on Cyg X-3 \citep*{Wat94,Col99,Cho02,Gal03,Szo08b}. 
It was found that in quiescence the soft X-ray emission ($<$ 12 keV) is correlated with the radio emission (GHz regime) and anti-correlated with the hard X-ray 
emission ($>$ 20 keV). During major radio flares, the hard X-ray emission correlates with the radio emission, while the soft X-ray emission is anti-correlated with 
the radio emission. Recently, based on the relationship between the radio and soft X-ray emissions, \cite{Szo08b} have revisited the classification of the radio states 
of \cite{Wal94,Wal95,Wal96}, identifying six radio/X-ray states for Cyg X-3: quiescent, minor flaring, suppressed, quenched, major flaring, and post-flare. 

\section{Observations and Analysis}

\subsection{New e-VLBI data}
Electronic very-long-baseline interferometry (e-VLBI) is a technique in which the data from radio telescopes separated by hundreds or thousands of km are streamed in 
real-time to the central data processor (i.e. the correlator). This technique is under continuous development and has been successfully used for 
the study of transient phenomena \citep{Rus07,Tud07}.

We observed Cyg X-3 with the European e-VLBI Network (e-EVN) at 5 GHz
on five occasions, in 2007 Jun and 2008 Apr (Table 1). The data
transfer rate was 256 Mbps for the run in 2007 and 512 Mbps for the
rest. During the observations the target did not vary significantly in flux density with the 
exception of the run on 2007 Jun 25 when a smooth gradient was recorded during the first two hours. Although efforts were made to assure an accurate imaging of the 
data, artifacts might be present in the radio map for this epoch
(Fig. 1). In particular, the reality of the extended radio emission
towards South-West is hard to assess with certainty. 
It appears also in the image made with a segment of the whole data set, when no variations in the amplitude took place and therefore it seems not to be an artifact. 

The data were calibrated in \textsc{AIPS} (e.g. \citealt{Gre03}) and imaged in \textsc{Difmap} \citep{She97} using standard procedures. We used, upon availability, 
3C 345 or 3C 84 as a fringe finder. 
J2007+4029, located 4\fdg7 away from Cyg X-3 was the phase referencing calibrator. To improve on the amplitude calibration of the data we used the
unresolved calibrator J2002+4725. Corrections of the antenna gains have been applied manually by 
imposing explicitly a constraint on the flux density of the mentioned point-like source, more exactly 0.5 Jy, as implied by the data.

\begin{table}
 \centering
  \caption{e-EVN observations of Cyg X-3 at 5 GHz. The table contains the date of the observation, the corresponding modified Julian Day, the total observing time and 
the participating radio telescopes (Cm-Cambridge, Mc-Medicina,
Jb2-Jodrell Bank Mk2, On-Onsala, Tr-Torun, Wb-Westerbork).}
  \begin{tabular}{@{}cccc}
  \hline \hline
 Date & MJD & Total time & Radio telescopes  \\
     &  day &   h        &                  \\
\hline
2007 Jun 25 & 54276  &  10.3  &  Cm Mc Jb2 On Tr Wb   \\
2008 Apr 09 & 54565  &  8.0   &  Cm Mc Jb2 On Tr Wb   \\
2008 Apr 23 & 54579  &  10.3  &     Mc Jb2    Tr Wb   \\ 
2008 Apr 25 & 54581  &  10.0  &  Cm Mc Jb2    Tr Wb   \\ 
2008 Apr 27 & 54583  &  9.1   &  Cm Mc Jb2    Tr Wb   \\ 
\hline \hline
\end{tabular}
\end{table}

\begin{center}
\begin{figure*}
  \includegraphics*[scale=0.37]{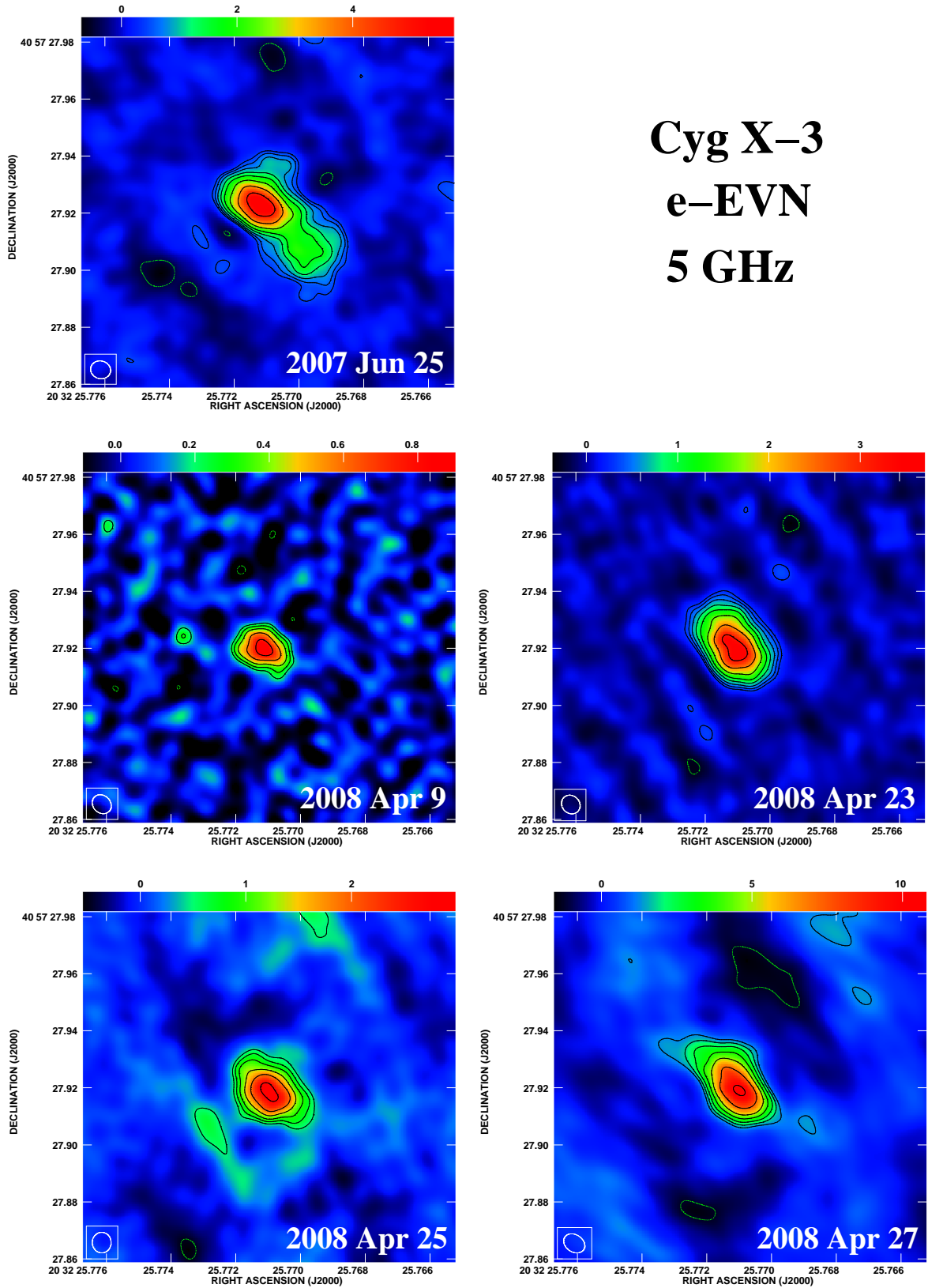}
  \caption{5 GHz e-EVN radio maps of Cyg X-3. The contours are at the
levels of -2.8, 2.8, 4, 5.6, 8, 11, 16, 23, 32, 45, 64, 90 times the
rms noises of respectively 0.15, 0.07, 0.10, 0.17 and 0.45 mJy beam$^{-1}$. The grey code bars on top of the maps are expressed in mJy beam$^{-1}$.}
\end{figure*}
\end{center}

\subsection{Archival VLBA and previous e-VLBI data}

In order to maximize the size of the sample to be used in the study we
complemented the recent e-VLBI observations with archival Very Long
Baseline Array (VLBA) and previous e-VLBI data. Whenever 
5 GHz data were not 
available, we used 15 GHz observations. This choice was motivated by the fact that this was the frequency closest to 5 GHz for which a relatively large amount of data 
was present in the archive. Whenever possible we used the results published in the literature (see the column ``comments'' in Table 2), however, some data have not 
been reported before or the information 
of interest to us (e.g. flux densities) was not available and in these cases we calibrated and imaged the data using standard procedures (see references in 
section 2.1). Before imaging, a selection in the {\it uv}-plane was made for the VLBA data, the details and reasons of which are presented in section 2.3. The calibrators 
3C 84 or 
J2202+4216 were used as fringe finders. Phase referencing was performed involving J2052+3635 or J2025+3343, situated respectively 5\fdg9 and 7\fdg4 away from Cyg X-3.
For four epochs, not listed in Table 2, three from 2000 Apr and one from 1995 May, it was not possible to identify with certainty the location of the target and hence 
were not considered in the study. 

\subsection{MERLIN data}

In support to the VLBI observations we also used Multi-Element Radio
Linked Interferometer Network (MERLIN) data for the
Cyg X-3 outburst of 2008 April (Fig. 2). MERLIN was operated for approximately
80 h between April 18--April 28 at 5 GHz with 6
radio telescopes: Jodrell Bank Mk2, Cambridge, Knockin, Darnhall,
Pickmere and Defford. The resulting maximum baseline length was 217 km,
i.e. about 3.5 M$\lambda$ at 5 GHz. The flux density scale was determined from observations
of 3C 286, using the three short baselines between the Jodrell Bank Mk2, Pickmere
and Darnhall radio telescopes. A flux density of 7.361 Jy was
assumed for 3C 286 \citep{Per86}. We used J2007+4029 as phase-referencing
calibrator. 

Only a small part of the flux density detected by MERLIN is recovered
at the VLBI scales. This is due to the lack of relatively short baselines in the
EVN array. 

\begin{center}
\begin{figure}
  \includegraphics*[angle=-90,scale=0.35]{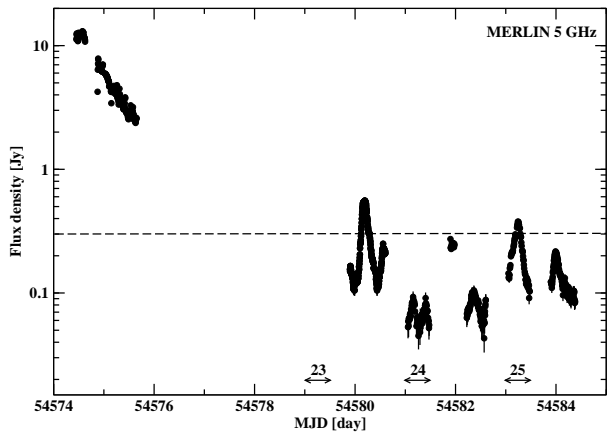}
  \caption{5 GHz MERLIN 5 min averaged radio light curve of Cyg X-3 between 2008 Apr 18-Apr
28. The time intervals of the e-EVN observations and their number (see
Table 2) are indicated. The horizontal dashed line corresponds to the transition level of 300 mJy (see section 3).}
\end{figure}
\end{center}

\subsection{Analysis of radio VLBI data}

\begin{table*}
 \centering
  \caption{Observational log. The table contains the number of the
observation, the corresponding date, the frequency of the radio
observation, the radio flux densities of the core and core+jet, the
ASM-RXTE count rate in the 3--5 keV energy band, the hardness ratio
HR2=(5--12 keV)/(3--5 keV), and comments.}
  \begin{tabular}{@{}cccccccl}
  \hline \hline
Obs & Date & Freq. & Core FD & Total FD & CR (ASM) & HR2 & Comments  \\
nr. &     &  GHz   &   mJy          &      mJy           &   c s$^{-1}$           &     &           \\
\hline
1  &  1997 Feb 06 (MJD 50485) & 15 &   224 $\pm$ 10 &   1800 $\pm$ 200  &   6.48 $\pm$ 0.09     & 1.59 $\pm$ 0.03   &     \\
2  &  1997 Feb 08 (MJD 50487) & 15 &   167 $\pm$ 5  &    537 $\pm$ 50   &   6.59 $\pm$ 0.10     & 1.50 $\pm$ 0.03   &     \\
3  &  1997 Feb 11 (MJD 50490) & 15 &   611 $\pm$ 5  &    611 $\pm$ 5    &   7.58 $\pm$ 0.10     & 1.41 $\pm$ 0.02   &     \\
4  &  2000 Apr 02 (MJD 51636) &  5 &  1500 $\pm$ 15 &   1500 $\pm$ 15   &  10.45 $\pm$ 0.33     & 1.09 $\pm$ 0.05   &     \\
5  &  2000 Apr 03 (MJD 51637) &  5 &   870 $\pm$ 6  &    870 $\pm$ 6    &  12.08 $\pm$ 0.31     & 1.17 $\pm$ 0.04   &     \\
6  &  2000 Apr 21 (MJD 51655) &  5 &         -      &   1890 $\pm$ 262  &   5.86 $\pm$ 0.13     & 1.86 $\pm$ 0.05   &     \\
7  &  2001 Sep 18 (MJD 52170) &  5 &    21 $\pm$ 5  &  13120 $\pm$ 300  &   6.96 $\pm$ 0.18     & 1.68 $\pm$ 0.06   & \cite{Mil04}    \\
8  &  2001 Sep 19 (MJD 52171) &  5 &         -      &  13120 $\pm$ 300  &   6.23 $\pm$ 0.13     & 1.89 $\pm$ 0.05   & \cite{Mil04}    \\
9  &  2001 Sep 20 (MJD 52172) &  5 &         -      &   6120 $\pm$ 300  &   5.11 $\pm$ 0.11     & 1.97 $\pm$ 0.05   & \cite{Mil04}    \\
10 &  2001 Sep 21 (MJD 52173) &  5 &    39 $\pm$ 5  &   4736 $\pm$ 300  &   6.22 $\pm$ 0.14     & 1.85 $\pm$ 0.05   & \cite{Mil04}    \\
11 &  2001 Sep 22 (MJD 52174) &  5 &    73 $\pm$ 5  &   2922 $\pm$ 300  &   4.96 $\pm$ 0.20     & 2.20 $\pm$ 0.10   & \cite{Mil04}    \\
12 &  2001 Sep 23 (MJD 52175) &  5 &     8 $\pm$ 2  &   1389 $\pm$ 300  &   6.34 $\pm$ 0.12     & 1.74 $\pm$ 0.04   & \cite{Mil04}    \\
13 &  2002 Feb 15 (MJD 52320) & 15 &    77 $\pm$ 6  &     77 $\pm$ 6    &   9.88 $\pm$ 0.26     & 1.00 $\pm$ 0.04   &     \\
14 &  2004 Jan 11 (MJD 53015) & 15 &   136 $\pm$ 4  &    136 $\pm$ 4    &   1.74 $\pm$ 0.16     & 2.65 $\pm$ 0.27   & ASM data from 2004 Jan 3  \\
15 &  2004 Oct 17 (MJD 53295) & 15 &    98 $\pm$ 3  &     98 $\pm$ 3    &   1.68 $\pm$ 0.09     & 2.29 $\pm$ 0.15   &     \\
16 &  2004 Nov 21 (MJD 53330) & 15 &    59 $\pm$ 2  &     59 $\pm$ 2    &   1.29 $\pm$ 0.09     & 3.10 $\pm$ 0.25   &     \\
17 &  2005 Jun 10 (MJD 53531) & 15 &    65 $\pm$ 1  &     65 $\pm$ 1    &   1.29 $\pm$ 0.12     & 2.56 $\pm$ 0.28   & ASM data from 2005 Jun 17  \\
18 &  2005 Dec 28 (MJD 53732) & 15 &    63 $\pm$ 2  &     63 $\pm$ 2    &   3.52 $\pm$ 0.34     & 2.04 $\pm$ 0.23   &     \\
19 &  2006 Apr 20 (MJD 53845) &  5 &    87 $\pm$ 9  &     87 $\pm$ 9    &   6.92 $\pm$ 0.15     & 1.40 $\pm$ 0.05   & e-EVN; \cite{Tud07}     \\                     
20 &  2006 May 18 (MJD 53873) &  5 &   101 $\pm$ 3  &    909 $\pm$ 18   &   5.79 $\pm$ 0.11     & 1.70 $\pm$ 0.04   & e-EVN; \cite{Tud07}    \\
21 &  2007 Jun 25 (MJD 54276) &  5 &    23 $\pm$ 1  &     42 $\pm$ 3    &   2.40 $\pm$ 0.12     & 2.24 $\pm$ 0.13   & e-EVN     \\
22 &  2008 Apr 09 (MJD 54565) &  5 &     4 $\pm$ 1  &      4 $\pm$ 1    &   8.91 $\pm$ 0.15     & 1.14 $\pm$ 0.03   & e-EVN     \\
23 &  2008 Apr 23 (MJD 54579) &  5 &    52 $\pm$ 4  &     52 $\pm$ 4    &   7.27 $\pm$ 0.45     & 1.65 $\pm$ 0.17   & e-EVN     \\
24 &  2008 Apr 25 (MJD 54581) &  5 &    19 $\pm$ 1  &     19 $\pm$ 1    &  10.83 $\pm$ 0.21     & 1.25 $\pm$ 0.04   & e-EVN     \\
25 &  2008 Apr 27 (MJD 54583) &  5 &    62 $\pm$ 1  &     62 $\pm$ 1    &   7.77 $\pm$ 0.14     & 1.31 $\pm$ 0.04   & e-EVN     \\
\hline \hline
\end{tabular}
\end{table*}

Cyg X-3 is heavily scattered \citep{Wil94} and this is of relevance for VLBI observations. The size of the scattering disc is \citep{Mio01}:
\begin{equation}
\theta = 448 \left(\frac{\nu}{\rm{GHz}}\right)^{-2.09} \qquad \mbox{mas}
\end{equation}
meaning about 15 mas at 5 GHz and 1.5 mas at 15 GHz.

The maximum {\it uv}-distance for which the data contain useful information is then given by:
\begin{equation}
uv_{max} = \frac{206.265}{448} \left(\frac{\nu}{\rm{GHz}}\right)^{2.09} \qquad \mbox{M$\lambda$}
\end{equation}
corresponding to 13 M$\lambda$ at 5 GHz and 132 M$\lambda$ at 15 GHz.

All the e-EVN observations reported here and labeled as such in column ``comments'' of Table 2 were observed at 5 GHz. Their maximum intrinsic {\it uv}-distance is 
about 24 M$\lambda$. A cut at 13 M$\lambda$ was found to be unnecessary with respect to improving the image quality and so the values given in Table 2 are 
based on maps with the original {\it uv}-distance range.
 
For the VLBA data the situation is different. Their maximum nominal {\it uv}-distance is around 140 M$\lambda$ at 5 GHz and close to 440 M$\lambda$ at 15 GHz. Imposing 
an upper cut in the {\it uv}-distance plane according to the limits given by equation (2) did result in a significant increase in the quality of the maps. We therefore 
adopted the procedure for all the VLBA data sets we analyzed. The only exceptions are the epochs from 2001 Sep when we quote the results from \cite{Mil04}. This is 
justified by the fact that these authors limit the maximum {\it uv}-distance taken into account in the imaging process by selecting a group of close-by antennae, which 
it turned out produced a very similar effect on the {\it uv}-plane to what our approach would. The two methods rendered insignificant differences in the resulting 
radio maps. 

The total flux densities (of core plus jets, if present) were measured in the image-plane using \textsc{Difmap} or \textsc{AIPS}. {\it uv}-plane fitting with elliptical 
Gaussians were also performed under \textsc{Difmap} as a check and the results were consistent, within the errors. 

The core flux densities are less secured, despite the quoted errors, due to unknown systematics. One of the largest source of uncertainty is the lack of knowledge 
of the precise proper motion of Cyg X-3. In this paper we adopt the value of the proper motion obtained by analyzing the VLBI data listed in Table 2. The detailed results 
are presented elsewhere \citep{Mil09b}.

In order to obtain the core flux density, whenever a jet was present in the image we subtracted it in the {\it uv}-plane and then measured the flux density at the 
expected position of the core after correcting for the proper motion. This approach is sensitive to any errors associated with the phase 
referencing process. Given the fact that for the VLBA observations the distance between the target and the phase reference calibrators was significant, such errors 
cannot be ruled out. In fact, in a few cases (flagged with a minus sign in the fourth column of Table 2) it was impossible to identify the core because of evident 
shifts from the expected position of Cyg X-3 which likely were due to unsuccessful phase referencing. However, in many of the images the target actually appeared 
as a single compact emitting region and we assumed it to be the core of the system, irrespective to any small shifts from the expected position as registered in a 
minority of cases. 

Correctly locating the core has far reaching consequences. Building their argument on the apparent lack of proper motion of what was considered to be the core in VLBA data
separated by 4.5 yr \citep{Mio01,Mil04}, \cite{Tud07} suggested that one of the compact emitting regions (denoted ``knot C'') in their e-EVN data (from 2006 May 18) 
is probably not the 
core of the system. However, in the light of the new knowledge provided by the analysis of the VLBI data with respect to the proper motion of the system, 
it is almost sure that ``knot C'' is actually the core. 
This has important implications with respect to the orientation of the
jet of Cyg X-3, but are beyond the scope of this work and are addressed
elsewhere \citep{Mil09b}. The core flux density
reported in Table 2 for the 2006 May 18 epoch is the flux density of ``knot C''. 

A clarification has to be made as well with respect to the short communication presented by \cite{Tud08} about the e-EVN run on 2008 Apr 9. The authors report that 
the preliminary analysis of the data reveals the presence in the radio map of two distinct compact emitting regions. After a detailed analysis we discovered that one 
of those emitting regions is an artifact. This was due to a temporary
bug in the standard post-correlation software used prior to the data release, which allowed for visibilities near scan boundaries to appear in
both scans, thus resulting in phase-reference source visibilities
in the Cyg X-3 data. This way an artificial source was introduced
near the phase-centre. Coincidentally, the location of the artifact was very close to the position 
of a compact emitting region (``knot A'') observed in the e-EVN
observations of 2006 May 18 making it even more believable. 
In Fig. 1 we present the correct version of the radio map corresponding to the 
2008 Apr 9 epoch. 

Note that the term ``core'' used throughout the paper does not
necessarily reflect the actual size of Cyg X-3 system (i.e. the size of
the region at whose boundaries the optical depth for the radio emission becomes unity). \cite{Mil09a} used the minimum variability 
timescales observed during a period of low-level activity to set
constraints on the size of the source at 43 GHz and 15 GHz. For an outflow traveling at the speed of light at a distance 
of 10 kpc it translated in a size of 0.2--0.3 mas. This is smaller then the size of the scattering disc at the two frequencies we used here. 

\subsection{X-ray data}

\begin{table*}
 \centering
  \caption{Classification of the PCA/HEXTE-RXTE X-ray spectra of Cyg X-3. The spectra were taken quasi-simultaneously with the radio data in Table 2. The table lists the 
number of the observation (associated to first column of Table 2), the date, the RXTE ID of the observation, the detectors used, and the X-ray spectral group following the 
nomenclature of \citealt{Szo08b}.}
  \begin{tabular}{@{}ccccc}
  \hline \hline
 Obs & Date & Obs ID & Detectors & Spectral  \\
nr. &     &   &  &  group        \\
\hline
5    &  2000 Apr 03 (MJD 51637) &  50062-02-01-00 & PCU 0,2; HEXTE A & 5 \\
6-1  &  2000 Apr 22 (MJD 51656) &  50062-01-01-00 & PCU 0,2; HEXTE A & 4 \\
6-2  &  2000 Apr 22 (MJD 51656) &  50062-01-01-01 & PCU 0,2; HEXTE A & 3 \\
20-1 &  2006 May 17 (MJD 53872) &  91090-03-01-00 & PCU 0,2; HEXTE B & 3/4 ? \\
20-2 &  2006 May 17 (MJD 53872) &  91090-03-02-00 & PCU 0,2; HEXTE B & 3/4 ? \\
23   &  2008 Apr 23 (MJD 54579) &  93434-01-01-00 & PCU 1,2; HEXTE B & 3/4 ? \\
24   &  2008 Apr 25 (MJD 54581) &  93434-01-02-00 & PCU 1,2; HEXTE B & 5 \\
25   &  2008 Apr 27 (MJD 54583) &  93434-01-03-00 & PCU 1,2; HEXTE B & 4 \\
\hline \hline
\end{tabular}
\end{table*}

\begin{center}
\begin{figure*}
  \includegraphics*[scale=0.35,angle=-90]{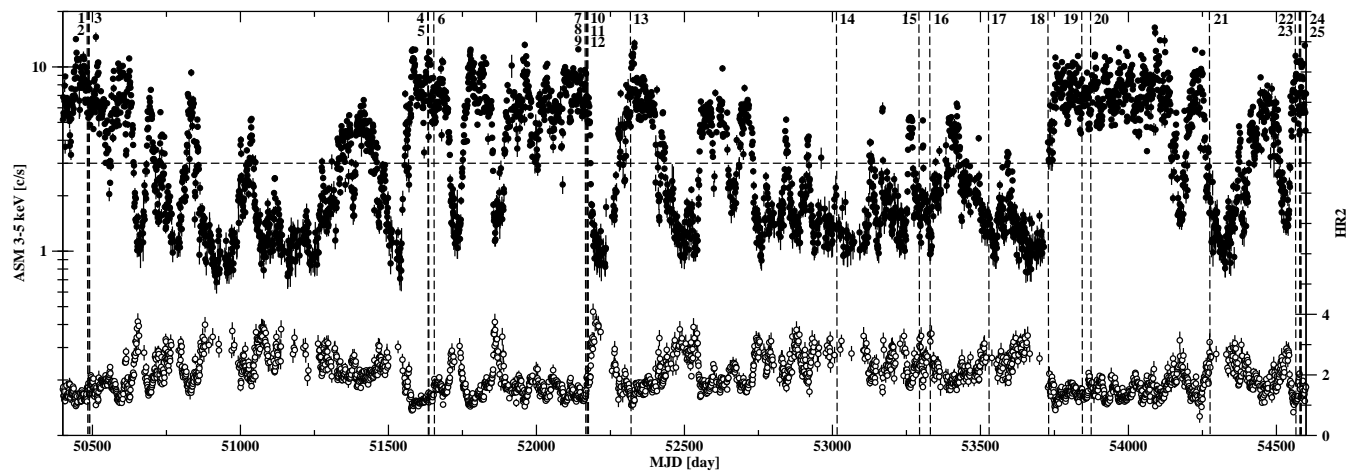}
  \caption{The daily-averaged ASM-RXTE 3--5 keV light curve of Cyg X-3
between 1996 Nov--2008 May (filled circles) and the ASM-RXTE hardness
ratio HR2=(5--12 keV)/(3--5 keV) for the 
same object and time interval (empty circles). The vertical dashed
lines correspond to the radio observations in Table 2, labeled by
observation number. The horizontal dashed line corresponds to the
transition level of 3 counts s$^{-1}$  (see section 3). }
\end{figure*}
\end{center}

We used Rossi X-ray Timing Explorer (RXTE) All Sky Monitor (ASM)/Proportional Counter Array (PCA)/High-Energy X-ray Transient Experiment (HEXTE) data observed 
quasi-simultaneously (i.e. within one day) with the radio observations. 

As pointed out for instance by \cite{Szo08b}, due to the shape of the X-ray spectra the best ASM band to study the radio/X-ray correlation in Cyg X-3 is the 3--5 keV and 
therefore further on, unless otherwise noted, we will refer to this band whenever an ASM count rate is quoted. To calculate the hardness ratio (HR2) we used the ASM count 
rates in the 5--12 keV and 3--5 keV bands. In order to put the particular values used in this work in retrospective, Fig. 3 shows the variations of the ASM count rate and 
hardness ratio between 1996 Nov--2008 May.

In the case of PCA we used the first layer only and the Proportional Counter Units (PCU) 0 and 2, or 1 and 2, upon availability. For HEXTE we analyzed data from either 
clusters A or B. The X-ray spectra were extracted in the PCA energy band 3--20 keV and HEXTE energy band 15--90 keV. The spectra were fitted in 
\textsc{XSPEC11} (e.g. \citealt{Arn96}) with a model 
including Comptonization by thermal and non-thermal electrons (\textsc{EQPAIR}; \citealt{Cop92,Cop99,Gie99}), Compton reflection (\textsc{PEXRIV}; \citealt{Mag95}), 
absorption by two neutral media, fully and partially covering the source (\textsc{ABSND}), and a broad Gaussian Fe K$\alpha$ fluorescent line (\textsc{GAUSS}). Similar 
models were used for Cyg X-3 by \cite{Vil03,Hja08,Szo08a,Szo08b}. A detailed description of the model is given by \cite{Hja09}. A systematic error of 3 per cent was added 
to the data. The fits had a reduced $\chi^2$ around 1 or less.

Following \cite{Szo08b} both in methodology and nomenclature (Fig. 4),
the resulting absorbed spectra were classified according to their shape
and flux at 20 keV 
(Table 3). 
In a few cases it was not possible to identify with certainty the X-ray spectral group.  

\begin{center}
\begin{figure*}
  \includegraphics*[scale=0.6]{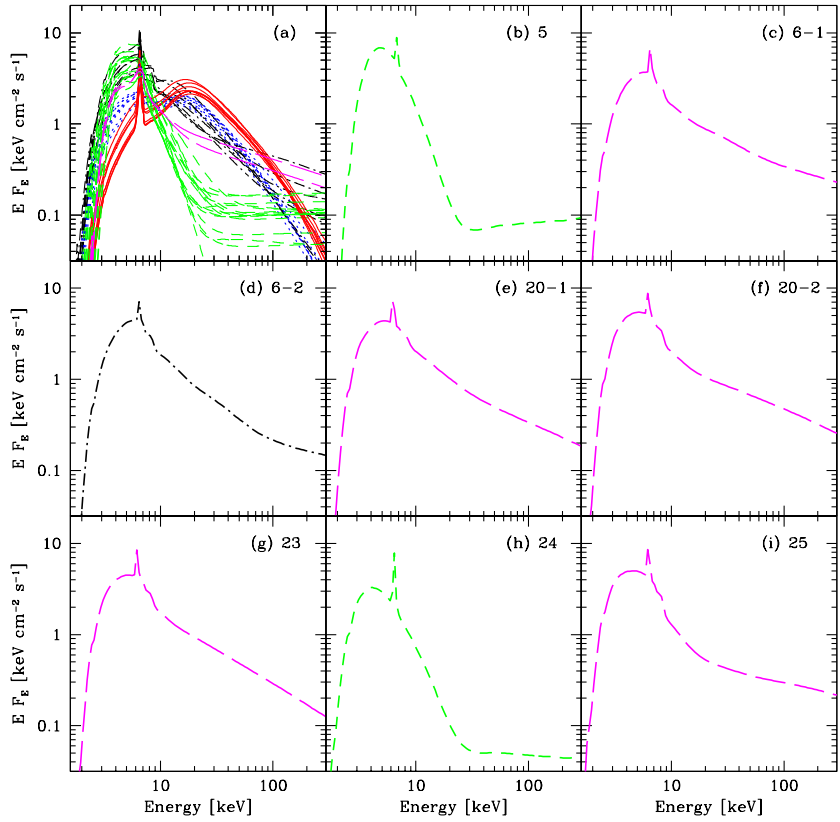}
  \caption{Classification of the X-ray spectra of Cyg X-3
\citep{Szo08b}. {\it Panel a):} PCA/HEXTE-RXTE spectra of Cyg
X-3 between 1996--2000. Group 1: continous red line; Group 2: dotted blue line; Group 3:
dash-dotted black line; Group 4: long-dashed magenta line; Group 5:
short-dashed green line. {\it Panels b)-i):} spectra corresponding to
observations in Tabel 3. Numbers indicate the observation epoch. For representation purposes the spectra for epochs 20-1, 20-2 and 23 were
considered to pertain to Group 4.}
\end{figure*}
\end{center}

\section{milliarcsecond behaviour}

Working with Green Bank Interferometer (GBI) radio data and ASM/PCA/HEXTE RXTE X-ray data between 1996--2000, \cite{Szo08b} classified six radio/X-ray states 
of Cyg X-3. Four of them correspond to the radio states previously identified by \cite{Wal94,Wal95,Wal96} using GBI observations obtained during the period 
1988--1992. Fig. 5, upper panel (an adaptation of Figs. 4 and 6 of \citealt{Szo08b}) shows the position of these states in a log-log representation of the radio versus 
X-ray flux, together with the corresponding X-ray spectral groups to which they can be associated. This schematic representation is purely phenomenological. It is 
based on the above cited works as well as on the VLBI data that will be discussed further on. The two transition levels, represented by the dashed lines, are chosen to 
coincide with the values favoured by \cite{Szo08b}: 3 counts s$^{-1}$
ASM count rate and 300 mJy radio flux density at 8.3 GHz.

Based on the monitoring campaigns with GBI (e.g. \citealt{Wal94}), Ryle Telescope (e.g. \citealt{Fen97,Poo06}), RATAN-600 (e.g. \citealt{Tru06,Tru08}), strong evidence 
is mounting that major 
flares of Cyg X-3 are preceded by periods of quenched radio emission. \cite{Szo08b} found that after a major flare state the system is entering the minor flaring or the 
suppressed states, via the post-flare stage. They also note that the minor flares happen during transitions, in both directions, between the quiescence and the 
suppressed states. The changes through different states are mirrored by variations in the X-ray spectra. An approximate correspondence can be established between them 
\citep{Szo08b}: spectral X-ray group 1 is associated to the the quiescence state, group 2 to the minor flaring state, group 3 to the suppressed state, group 4 to the 
major flaring state, group 5 to the post-flaring state. This was a summary of a few of the characteristics of Cyg X-3 when the arcsec scale radio observations are 
considered. However, at VLBI scales (i.e. mas scales) extended radio emission (i.e. jets) was resolved \citep{Mio01,Mil04,Tud07}. Unfortunately, 
such high angular resolution radio observations are much more sparse
than those at lower angular scales and hence it is not yet clear, for
instance, what is the interplay between 
the radio emission from the core and from the jets, or what is the exact nature of the jets themselves (discrete knots of adiabatically expanding plasma or internal 
shocks propagating within a quasi-steady flow). 

In this context, despite the limited amount of data available at the moment, an attempt at probing the behaviour of 
Cyg X-3 at mas scales has two important advantages over the arcsec scales approach: it potentially offers a more accurate image of the system, free of possible 
flux contamination from other sources present in the field of view, and it gives the opportunity to disentangle the jet contribution to the total flux density. 

\begin{center}
\begin{figure}
  \includegraphics*[scale=0.33]{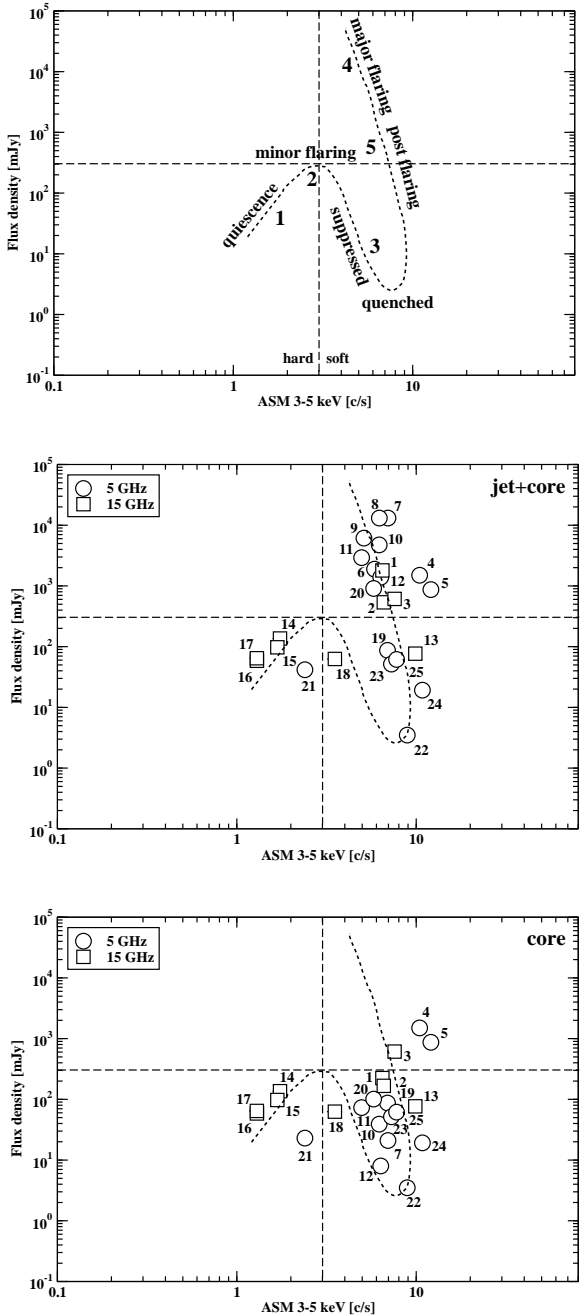}
  \caption{{\it Top}: The radio/X-ray states of Cyg X-3 as proposed by
\citealt{Szo08b}. The numbers correspond to the X-ray spectral
group. The transition levels at 3 
counts s$^{-1}$ and 300 mJy are chosen to coincide with those used in
the above cited reference. The curved dashed line is a guide for the
eye and mirrors the trend 
found by 
those authors. {\it Middle}: Quasi-simultaneous radio/X-ray emission of Cyg X-3 when the radio emission of core and jet are both considered (data from Table 2). 
Radio data at 5 GHz 
represented with open circles and at 15 GHz with open squares. The numbers correspond to the observation number (first column of Table 2). {\it Bottom}: The same as 
middle panel, but this time only the radio emission from the core of the system is taken into account.}
\end{figure}
\end{center}

Fig. 5, middle panel, shows the relation between the soft X-ray and radio emissions when both the contribution of the jet and core are taken into account in the radio 
band. In the same figure, in the bottom panel, the contribution of the jet to the radio flux density has been removed. As mentioned at the beginning of this section, 
the superimposed dashed lines should be taken as guides for the eye and reflect the model of the radio/X-ray states of Cyg X-3 as proposed by \cite{Szo08b}. In defining 
this scheme these authors used GBI data at 8.3 GHz, but noted that the use of 2.25 GHz GBI data or the 15 GHz Ryle Telescope data yielded relatively similar results. 
This gives us confidence that the use of the two frequencies in our study, namely 5 GHz and 15 GHz, on the one hand should not infringe on the direct comparison with 
their results, and on the other should not affect significantly any trend present in the data due to the unaccounted for effect of the shape of the radio spectrum 
at the moment of the observations. Regarding this latest aspect we also add that after the onset of an outburst period, that is when the system is in the major flaring, 
post-flaring or minor flaring state, the radio spectrum between 5 and 15 GHz is usually optically thin \citep{Mil04,Tru06,Lin07,Tru08,Mil09a} which means that the 15 GHz 
flux densities from Table 2 and Fig. 5 corresponding to these three states are likely lower limits for the unknown 5 GHz flux densities. But again, given the log-log 
representation of Fig. 5, this should not affect any of the conclusions to follow. 

With respect to Fig. 5, middle panel, acknowledging the small number of observations, the mas scale behaviour of Cyg X-3 does seem consistent with the arcsec scale one 
reported by \cite{Szo08b} in their Figs. 4 and 5, at least from the point of view of the overall distribution of points and the normalization. 

\subsection{Quiescence state}

The observations 14, 15, 
16 and 17 strongly suggest that the system was in the quiescence state: the levels of the radio and X-ray emissions as well as the apparent correlation between them 
are those expected, but note that the X-ray data is not simultaneous with the radio on two of the epochs. 

The observation 21 (2007 Jun 25) requires some discussion. 
About three weeks before, Cyg X-3 exhibited a major flare \citep{Tru07}
with the peak flux density on 2007 Jun 1. Within one week the flux
density decreased 
significantly, 
from $\sim$3.5 Jy down to $\sim$200 mJy and the monitoring stopped. No further indication with respect to the immediate radio evolution exists. However, one might make 
the case that during the observation 21 the system was likely in quiescence. It might be that after the major flare faded away, Cyg X-3 passing through the post-flare 
state entered the minor flaring state (via the suppressed state perhaps) and from there ended up in the quiescence. Then the jet observed on 2007 Jun 25 (Fig. 1), if 
real (see section 2.1), is the reminiscence of the excursion in the minor flaring state. Support for this scenario comes from the ASM X-ray data. After the major flare 
about 2007 Jun 1 the X-ray emission became harder, with count rates decreasing steadily from around 12 counts s$^{-1}$  down to the transition level of 3 counts s$^{-1}$ 
about 10 days later. The system crossed the threshold and after reaching 2 counts s$^{-1}$ evolved back towards softer X-ray emission, crossed again the transition level 
and, finally, made a last pass over the 3 counts s$^{-1}$ limit (close to 2007 Jun 20), ending up in the region of $\sim$2 counts s$^{-1}$ (and going dimmer) where it was 
probed by the radio observation of 2007 Jun 25. 

\subsection{Suppressed state}

During the observation 18 Cyg X-3 was very likely in a suppressed state
(see Fig. 4 of \citealt{Tru06}). This state, introduced by
\cite{Szo08b}, is characterized by the fact that it is not immediately
followed by a major radio flare. 

The isolated observation 13 on 2002 Feb 15 is quite hard to classify with certainty. The only independent radio information available is from the unpublished 
Ryle Telescope data on G.G. Pooley's web page\footnote{http://www.mrao.cam.ac.uk/$\sim$guy/cx3/2002.ps.gz}: the system was in a kind of suppressed state, just a few 
days before a quenched state. 

\subsection{Quenched state}

On 2008 Apr 9 (observation 22; Fig. 1) Cyg X-3 was at the level of 4
mJy at 5 GHz. The system was in a quenched radio state. This is the
faintest radio VLBI detection of Cyg X-3 at cm wavelengths so far. Two
weeks later the system was already in an active state (observations 23,
24, 25). Such a behaviour is
consistent with previous observations showing that major radio flares of Cyg
X-3 are preceded by periods of quenched radio emission.   

\subsection{Major flare/post flare states}

Two weeks after the faint radio detection on 2008 Apr 9, a series of
three e-EVN observations (23, 24 and 25; Fig. 1), separated by 2 days
followed. Meanwhile the system underwent a major 
outburst with the peak on 2008 Apr 18 (see Fig. 2) and these observations 
actually reveal Cyg X-3 during the descending stage of the flare (however, 
note the significant rebrightening on 2008 Apr 27). X-ray spectra were also available on these dates (Table 3). For observation 23 it is not clear to which group 
the spectrum belongs too; it might be group 3 or 4. Given the fact that
the next two observations show spectra pertaining to groups 5 and 4
respectively, the likely interpretation is that the system didn't have
time to reach the suppressed state by the time of observation 23, but
was rather still in the major flaring/post-flaring states. 

On 2006 Apr 20 and 2006 May 18 (observations 19 and 20; Fig. 2 of \citealt{Tud07}) Cyg X-3 was observed a few weeks and a few days respectively after the peak of two 
major flares. In the first instance the system was in a post-flare state, while in the second very likely still in the flare state. The X-ray spectrum taken on 2006 
May 17 (Table 3) does not permit a clear classification, but given the radio data (e.g. the presence of a resolved jet) it would be perhaps reasonable to expect that 
the spectrum belongs to group 4. 

The observations 1 to 3 started two days after a major flare and were made when the system was in the major flare/post-flare states (Fig. 1 of \citealt{Mio01}). Much like 
during the e-EVN observations of 2008 Apr, a rebrightening event was detected on 1997 Feb 11, which can very likely be associated with the core of the system. 

Some very interesting data were taken in 2000 Apr (observations 4, 5 and 6). No reference was found in the literature to these runs. According to unpublished Ryle 
Telescope data\footnote{http://www.mrao.cam.ac.uk/$\sim$guy/cx3/2000.ps.gz}, the observations were made within a couple of days after two major outbursts. During 
observation 4 the system was in a major flare/post-flare state in which it persisted also during the observation 5, as suggested also by the X-ray spectrum (Table 3) 
with characteristics particular to group 5. Regarding observation 6, judging by the radio data alone, Cyg X-3 seemed to have been in a major flare/post-flare state. 
However, the X-ray data (Table 3) is intriguing since it suggests that the spectra changed during the run from group 4 to group 3, i.e. the system switched from the 
flare state to the suppressed state. In the context of the model of \cite{Szo08b}, such an evolution is to be expected but not at this quite 
high radio flux density. Recalling the situation of the observation 20 for which the X-ray spectrum revealed ambiguity with respect to the exact classification we can 
note that in both cases, i.e. observations 6 and 20, the radio emission can be disentangled in a core component and a jet component (although for observation 6 it was 
impossible to identify with certainty the location of the core; see section 2.3). The high radio flux density in these cases is due to the jet component, 
while, perhaps, in fact the core of the system was already close to the suppressed state (see the location of observation 20 in Fig. 5, bottom panel, after removing the 
jet component). Anyway, this implies that very quick spectral changes are possible in the core of the system. 

A very good coverage of the evolution of a major flare was obtained in 2001 Sep (observations 7--12; Fig. 1 of \citealt{Mil04}). Cyg X-3 was observed at or slightly after 
the peak of the outburst and during the dimming phase. It is no doubt that these data correspond to the major flare state of the system. 

\subsection{Jet/core disentanglement}

Insofar, the focus of the discussion was on the mas behaviour of Cyg X-3 when the radio emission from both the core and jet were taken into account. Removing this later 
component of the radio emission, we obtained the plot represented in Fig. 5, bottom panel. The trend in the distribution of the points observed at mas scale in Fig. 5 
middle panel and which looks so similar to that observed at arcsec scales by \cite{Szo08b} is lost. Basically what happens is that the data in Fig. 5 corresponding to the 
major flare/post-flare states, in which the jet emission is dominant, are shifted towards lower radio flux densities thus blurring the classification of the states, 
roughly speaking, beyond and below the ad-hoc X-ray and respectively
radio transition levels. While these states may be able to be
classified via their X-ray spectra, the radio emission cannot be used
as a diagnostic here. 

The important conclusion is that since in active
states most of the radio emission is
not coming from the core then during outbursts the overall radio flux is not a direct
tracer of the accretion state. 

However, when the overall radio flux is considered we do observe an anti-correlation/trend between the radio and X-ray
emissions in the flare/post-flare states (Fig. 5, middle and Fig. 4 of
\citealt{Szo08b}). Moreover, also in the soft X-ray states of the system
(more precisely on the major flare branch and part of the post-flare
branch) \cite{Szo08b} found a correlation/trend (their Fig. 7) between
the radio emission and the hard X-rays in the energy band 20--100 keV as measured by the Burst and
Transient Source Experiment (BATSE) on-board the Compton Gamma-Ray
Observatory. These correlations or, more conservatively, trends are
quite interesting and deserve an attempt at explanation. Using arcsec
scale radio data \cite{Lin07}
and \cite{Mil09a} fitted satisfactorily the light curves of Cyg
X-3 during minor as well as major outbursts. They employed a
shock-in-jet model originally developed by \cite{Mar85} and later
generalized by \cite{Tur00}. The model assumes an adiabatically
expanding, conical, constant Doppler factor jet through which a shock
is propagating. Stronger radio flares seem to evolve on longer timescales than
the fainter ones, consistent with the formation of the shocks
downstream in the jet, further away from the core of the system. With
this in mind, the radio/X-ray trends mentioned above can be qualitatively
explained within the truncated disc model of the X-ray states
(\citealt*{Rem06,Don07} for reviews). In the soft X-ray states the
innermost region of the accretion disk is relatively close to the compact
object and slowly recedes further away as the outburst proceeds. As the
disc gets colder the level of the soft X-rays (i.e. ASM and PCA energy bands) is decreasing, while the
corona starts to build up leading to an increase in the hard X-rays
(i.e. HEXTE and BATSE energy bands). Therefore stronger radio flares,
which tend to peak at later times, will be associated with states in
which the accretion disc is truncated further away from the compact
object, where the soft X-ray levels are lower and the hard X-ray
levels are higher. This explanation is reasonable, but it should be
noted that other factors are likely to be involved in the behaviour of
Cyg X-3 during major outbursts. For instance the interaction of the jet
with the interstellar medium or the stellar wind from the companion is
to be expected during major flares. The external shocks thus formed
will perhaps contribute a fair amount of radio synchrotron radiation
to the overall radio emission. Unfortunately the environment of Cyg X-3
is poorly known, and given also the orientation of the jets in the system,
supposedly close to the line of sight, make any such estimations uncertain.  

Continuing the discussion on the multi-wavelength behaviour, in some XRB systems in hard X-ray states correlations seem to exist between the
radio and X-ray emissions.  Also, above some X-ray flux, in
the soft X-ray states, the radio emission
drops significantly. These properties are relatively well established in the case of black hole XRBs
(BHXRBs) [\citealt*{Cor03,Cor08,Gal03,Gal06}] and very tentative in the case
of neutron star XRBs (NSXRBs) [\citealt*{Mig03,Mig06,Tud09}]. A similar
behaviour is observed as well in Cyg X-3, as can be seen in Fig. 4 of
\cite{Szo08b} and Fig. 5 of the present work. Unfortunately, this does not offer
important hints with respect to the nature of the compact object in the
system. 

Looking forward to the future, an increase in the number of high 
resolution observations 
should offer more insights into the complex behaviour of Cyg X-3. But the quantity of observations in itself
is not a guarantee of advancement. Ideally one would have to probe the
system in different stages of its 
evolution. This is certainly doable. Our VLBI detection of the system
in a relatively deep quenched state (2008 Apr 9) shows that Cyg X-3 can be traced over a very broad range of flux 
densities. So far the VLBI observations have been performed almost exclusively in response to major outbursts and if they happen to catch Cyg X-3 in a different state 
it was mainly due to the slow reaction time of the radio facilities
after the triggering. With proper planning, it is possible to trigger
on the pre-flare quenched state to observe a major flare at high
resolution right from the onset to the end.  The e-VLBI technique, with
its rapid turnaround time would then allow to optimize the response to
such an outburst by modifying the observing strategy in real time as
necessary to best track the development of the flare. Furthermore, it
can offer the practical possibility to observe more ephemeral states (like perhaps the post-flare) in which the system spends only days to weeks. 

\section{Conclusions}

We have reported new e-VLBI radio observations of Cyg X-3 and analyzed
them together with previous e-EVN and archival VLBA data. Support MERLIN
observations were also presented. We have complemented 
the radio observations with quasi-simultaneous X-ray data: ASM-RXTE, and in a few cases pointed PCA/HEXTE-RXTE. 

It was found that the behaviour of Cyg X-3 at mas scales, as probed here, is well described by the radio/X-ray classification scheme proposed by \cite{Szo08b} based 
on arcsec scale radio observations, when the whole contribution of the
system to the radio flux density is taken into account (i.e. radio
emission from both jet and core). 
More precisely, this means that our results are in good agreement with those obtained by \cite{Szo08b} from the point of view of the evolutionary track followed by 
the system in the radio flux density/X-ray count rate plane as well as from the point of view of the overall normalization of this pathway. Equally important, we didn't 
find any clear evidence for departures in the behaviour of Cyg X-3 from that expected within the above mentioned model. Some minor question marks were indeed raised 
by a few observations, but given the small data set available no confident conclusion can be inferred with respect to their origin, statistic or not. 

However, when the contribution of the jet to the total radio flux density is removed, and therefore only the radio emission corresponding to the core is 
considered, the situation changes significantly. What is obvious is that when Cyg X-3 is in an active state (for sure during the major flaring and 
most of the post-flaring states) the observed radio flux density at cm
wavelengths is dominated by the emission from the jet. Hence the data 
imply that during these states there is no unambiguous connection
between the accretion state and the total radio emission.   

The mas behaviour of Cyg X-3 during major outbursts seems consistent
with the shock-in-jet model \citep{Lin07,Mil09a} but the interaction of
the jet 
with the surrounding medium should also play a role although its
importance is not known for the moment. Hence the more intimate relation 
between the radio emission of the core and the jet with respect to the radio/X-ray states of the system is not evident from the observations analyzed here. At least 
partially this is due to the limited size of our data set: only on a few epochs a jet component was present and its contribution to the radio flux quantified. Then the 
systematic errors associated to the identification of the location of the core might play an important role in altering any such relation. But the prospects for future 
are not at all bleak. We have shown that the VLBI technique can be a powerful tool in probing the behaviour of Cyg X-3. Carefully chosen observational set-ups 
(particularly with respect to the selection of the phase referencing calibrator) and, importantly, a better knowledge of the proper motion of the system, will no doubt 
alleviate some of the difficulties we encountered in the present analysis. 

\section*{Acknowledgments}

The European VLBI Network (EVN) is a joint facility of European, Chinese, South African and other radio astronomy institutes funded by their national research councils. 
e-VLBI developments in Europe are supported by the EC DG-INFSO funded Communication Network Developments project ``EXPReS'', Contract No. 02662. The National Radio 
Astronomy Observatory (NRAO) is a facility of the National Science Foundation operated under cooperative agreement by Associated Universities, Inc. The X-ray data were 
provided by the ASM/RXTE teams at MIT and at the RXTE SOF and GOF at NASA's GSFC. AS is supported by the European Community via contract ERC-StG-200911.

\end{document}